# Dynamic Estimation of Power System Inertia Distribution Using Synchrophasor Measurements

Mingjian Tuo
*Student Member, IEEE*
Department of Electrical and Computer Engineering
University of Houston
Houston, TX, USA
mtuo@uh.edu

Xingpeng Li
*Member, IEEE*
Department of Electrical and Computer Engineering
University of Houston
Houston, TX, USA
xli82@uh.edu

***Abstract*— Integration of intermittent renewable energy sources in modern power systems is increasing very fast. Replacement of synchronous generators with zero-to-low variable renewables substantially decreases the system inertia. In a large system, inertia distribution may vary significantly, areas of low inertia are more susceptible to frequency deviation, posing risks of load shedding and generation trip. Therefore, it is necessary for operators to evaluate and quantify the system inertia and its distribution in real time. This paper proposes a novel synchronized phasor measurement units (PMUs)-based dynamic system inertia estimation method. The proposed inertia estimation method is derived using electrical distance and clustering algorithm, which considers the impact of location of measurements relative to in-feed load and impact of oscillations. The center of inertia (COI) area and area of low inertia are also determined during the estimation. Numerical simulations are conducted on the IEEE 24-bus system with various load profiles using Transient Security Analysis Tools (TSAT), a core module of the DSATools, which demonstrate the efficacy of the proposed approach.**

## I. INTRODUCTION

System frequency deviation indicates the degree of imbalance between the system generation and consumption. Since it is not possible to maintain a perfect power balance at a nominal frequency at every instant, the unbalance between generation and consumption exists all the time. For instance, if the load exceeds the generation, energy must be drawn out of the large rotating masses of synchronized generators and fed into the power grid to compensate the mismatch of power balance, resulting in the decrease in generator rotational speed and system frequency. As long as the frequency fluctuates within the system operational limits, it would not cause issues for the grid. Under normal conditions, the frequency in ERCOT varies between 59.97 and 60.03 Hz [1].

Traditionally, conventional synchronous generators play a crucial role in system frequency regulation; the inertia provided by synchronous generators has important positive effects on system stability. However, with increasing volume of renewable energy sources, synchronous generators are gradually being replaced by renewable energy sources that have low to zero contribution to the system inertia. The total global installed capacity has increased by a factor of about 6 for wind power and a factor of 40 for solar power in the past decade [2]. In Australia, the level of combined wind and solar capacity has reached 20% in the National Electricity Market [3]. For the Nordics, nuclear power plants have been replaced by renewable generation, low inertia is listed as one of the three main challenges faced by the system [4]. In the ERCOT system, wind power generation contributes to 20% of the total generation capacity and provides around 15% of the total electric energy consumption on average [5].

Due to the degradation of system frequency response [6], conventional methods are not fast enough to halt a frequency deviation. Frequency regulation becomes much more important for the future low inertia power systems; in addition, it becomes more difficult to determine the regulation reserve requirement. To address this challenge, many frequency control schematics have been developed. The synthetic governor control method reserves the wind power generation by working in the over-speed zone instead of maximum power point tracking (MPPT) [7]-[8]. Wind power plant inertia control takes advantage of the kinetic energy stored in wind blades and turbines and provides a synthetic inertial frequency response in seconds [9]. The virtual inertia method presented in [10] imitates the kinetic inertia of synchronous generator to improve the system dynamic behavior.

The introduction of Wide Area Measurement System (WAMS), which utilizes phasor measurement units (PMUs), leads to online power system monitoring and analysis. Researchers have proposed many inertia estimation methods based on power system dynamic behaviors. [11] uses PMUs to obtain high resolution measurements to estimate the effective system inertia. Paper [12] develops a method which evaluates the demand side contributions to system inertia based on recorded measurements of frequency outage events. [13]-[14] created indices and methods to estimate system inertia distribution over the grid. Studies have shown that disturbances take some time to propagate through the whole power system and that frequency in an area of low inertia shows large deviation relative to other areas of high inertia. System inertia estimation methods based on the rate of change of frequency (RoCoF) measurements may suffer high noise and bias. [13] and [14] didn't discuss thoroughly about the impact of measurement location relative to perturbations on system inertia estimation.

To bridge the gap presented above, this paper proposed a PMU measurements-driven method which estimates the dynamic system inertia distribution and determines the center of inertia (COI) area. The frequency response under different renewable generation penetration levels is first tested. Then, an index based on electrical distance is used to estimate the inertia distribution over the entire grid. Butterworth filter is introduced in this paper to mitigate the impact of noise-induced measurement errors. To reduce the bias from location of measurements relative to the location of in-feed loss, disturbances on different buses over an observation window are combined; then a clustering algorithm based on electrical distance is utilized to accurately estimate the location of COI suitable for measurements. Areas with different inertia distribution levels are proposed to provide useful information to generation dispatch and frequency control.

The remainder of this paper is organized as follows. In Section II, the specifics of system identification and extraction of inertia values is described. Section III details the method of dynamic inertia distribution estimation and total system inertia calculation. Section IV presents the simulation results on the IEEE 24-bus test system. Section V presents the concluding remarks.

## II. Frequency Response and Scnario Development

### A. *Power System Inertia*

When a disturbance occurs, energy in the system should be redistributed to compensate the unbalance between power production and consumption. For a single rotating machine, the nominal inertia of it is equal to its kinetic energy $E_{rotation}$ in megawatt seconds (MWs) at rated speed, which is determined by the moment of inertia and rotational speed.

$$E_{rotation} = \frac{1}{2} J_i {\omega_n}^2 \tag{1}$$

where $J_i$ is the moment of inertia of the shaft in kg·m$^2$s and $\omega_n$ is the nominal speed in rad/s.

As shown in (1), the nominal inertia provided by a single generator is not related to the actual output power of a generator. The inertia of a single rotating shaft is commonly measured by its inertia constant, which is the per-unit value of inertia depending on the base value of the rated apparent power. For a single machine, the inertia constant can be express as:

$$H_i = \frac{J_i {\omega_n}^2}{2 S_{B_i}} \tag{2}$$

where $H_i$ is the inertia constant of the machine in seconds, $S_{B_i}$ is the base power in MVA.

A power system that connects multiple generators can be considered to act as a single equivalent center of inertia, the nominal value of the total system inertia $E_{sys}$ is the summation of the kinetic energy stored in all rotating machines synchronized with the grid. It can be expressed in the form of both the stored kinetic energy and inertia constants.

$$E_{sys} = \sum_{i=1}^{N} \frac{1}{2} J_i {\omega_n}^2 = \sum_{i=1}^{N} H_i S_{B_i} \tag{3}$$

The total inertia constant of a power system is given by

$$H_{tot} = \frac{\sum_{i=1}^{N} H_i S_{B_i}}{\sum_{i=1}^{N} S_{B_i}} \tag{4}$$

where $S_B$ denotes the total power rating of the system and is defined in (5).

$$S_B = \sum_{i=1}^{N} S_{B_i} \tag{5}$$

### B. *Frequency Response*

An objective of system operations is to ensure electricity production and consumption matched. However, it is not possible to maintain a perfect power balance, which in turn leads to the deviation in system frequency. For normal load fluctuation, the system frequency would not deviate beyond the nominal range; however, for large power imbalance caused by sudden loss of a generator or step wise load increase, the system frequency may deviate far away from the nominal range.

The dynamic behavior of the system frequency during a short period of time following a power mismatch event can be represented by generator swing equation. Given a generator $i$, the swing equation can be expressed as

$$\frac{df_i}{dt} = \frac{P_{iM} - P_{iE}}{2 H_i S_{Bi}} f_0 = \frac{\Delta P_i}{2 H_i S_{Bi}} f_0 \tag{6}$$

where $P_{iM}$ is the output mechanical power of the machine, $P_{iE}$ is its electrical load power, $f_0$ is the system frequency at the time of disturbance, $f_i$ is the electrical frequency, $\mathrm{d}f_i/\mathrm{d}t$ is known as the rate of change of frequency (RoCoF).

As an approximation, an equivalent equation can be applied to the whole system. Following a power mismatch event, the swing equation relating the RoCoF to the total system inertia is defined in (8).

$$\frac{\mathrm{d}f}{\mathrm{d}t} = \frac{-\Delta P}{2 H_{tot} S_B} f_0 \tag{7}$$

where $\Delta P$ is the change in system active power in MW.

The characteristics of power systems are very complex due to the existence of multiple electromechanical oscillation modes, system control noise, and variant distribution of inertia throughout the grid. The principal frequency dynamics can be described by the evolution of the center of inertia (COI) speed $f_{COI}$ [15]-[16], which is defined as

$$f_{COI} = \frac{\sum_{i=1}^{N} H_i f_i}{\sum_{i=1}^{N} H_i} \tag{8}$$

where $N$ is the total number of synchronous generators, $H_i$ is the $i^{\text{th}}$ unit's inertia constant, and $f_i$ is the angular frequency of the rotor of the $i^{\text{th}}$ generator. In this way, a system could be considered as a single equivalent center of inertia.

### C. *Model Development*

Renewable energy resources have been recognized as the most promising low carbon generations. In recent years, wind and photovoltaic (PV) power plants have witnessed a significant growth. Giga Watts (GW) wind and PV generation have been installed in many countries. For some countries in Europe, the wind or PV generation may even be able to meet most of electricity demand [7].

The Texas Interconnection (TI) is one of the three interconnection power systems in the U.S. Electric utilities in TI are electrically tied together during normal system conditions and operate at the same synchronized frequency around 60 Hz. Since wind generation has a larger share than PV in TI. In our modeling, double fed induction model "WGNC" in Transient Security Analysis Tools (TSAT) template is used as renewable generator model. In order to enhance the model credibility based on the scenarios developed, generators governor control model and exciter control model are considered. To validate the simulation model we build and evaluate the impacts of wind generation on system frequency responses, different penetration levels of renewable generation are modeled. The penetration rates are chosen to be 0%, 10%, 20%, 30% and 40%, respectively.

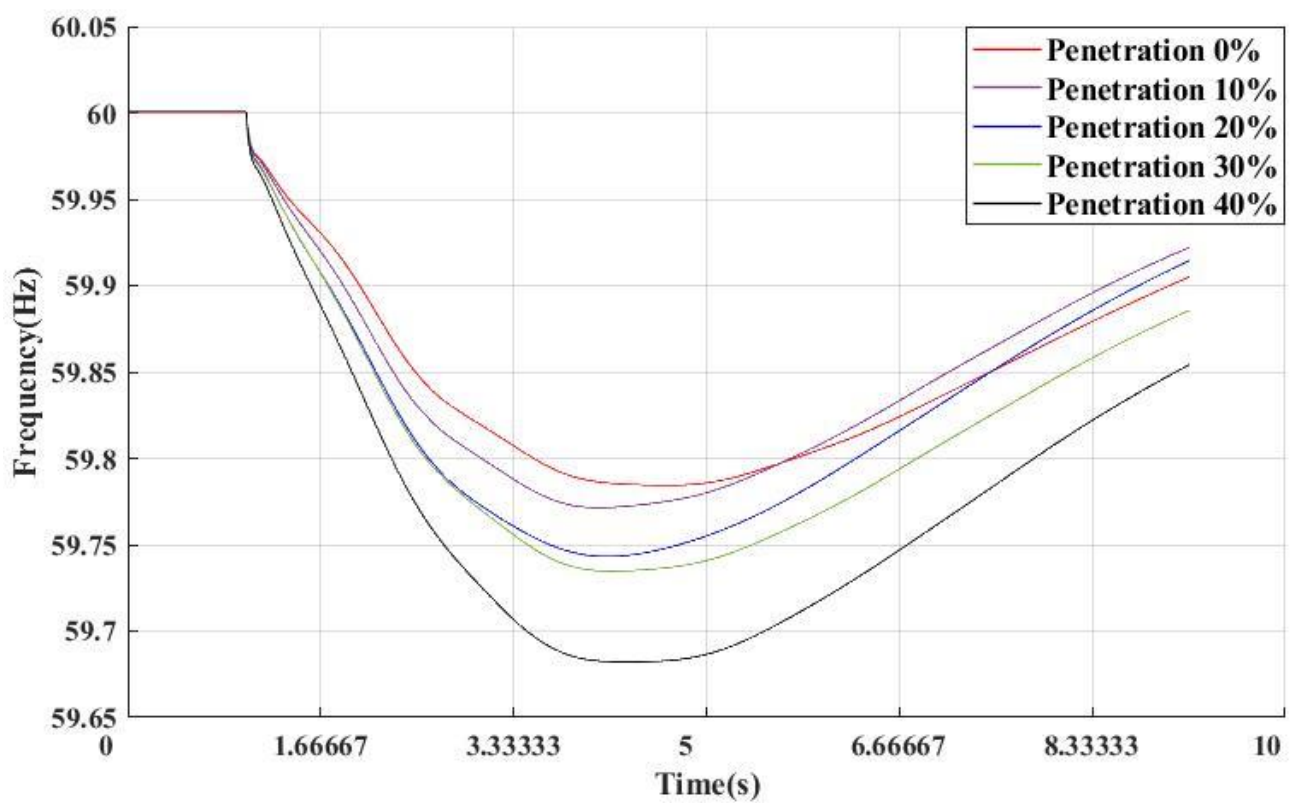

Fig. 1. Frequency response under different penetration levels of renewable generation on the IEEE 24-bus system.

The frequency response after the trip of 200 MW generation at bus 23 at $t$=1s is presented in Fig. 1. It can be observed that the system frequency response declines dramatically as the renewable generation penetration rate increases. The RoCoF for the scenario of 40% renewable penetration is 0.164 Hz/s, which is significantly larger than the RoCoF (0.113 Hz/s) for the scenario with no renewable generation; and the nadir drops by 0.1 Hz from 59.78 Hz to 59.68 Hz. The results indicate that for systems with higher RES penetration level, accurate inertia distribution and quantity estimation, proper inertia requirement, and additional countermeasures are required to address the issue of declining frequency response.

## III. METHODOLOGY

### *A. Measurement Preprocess*

The technique for inertia estimation requires accurate time-synchronized PMU data [11], [13]. It has been proved [16] that a method of curve fitting is required to mitigate the impact of measured transients in frequency following a loss, otherwise the calculated RoCoF may be significantly larger than the true value. Governor and exciter control may also introduce noises into the system resulting in frequency distortion. Fig. 2 shows the system frequency as measured from 3 PMUs installed on bus 7, 14 and 22 respectively, in response to an in-feed load of 31.15 MW on bus 8. It can be observed that the frequency and RoCoF measured on bus 7 and bus 22 show a significant deviation.

It is proved that a low pass Butterworth filter with a 0.5 Hz corner frequency can isolate the dominant system frequency from high frequency noise and improve the accuracy of measurements. However, as shown in Fig. 3, the filtered frequency signal indicates that the method is unsuitable when there are oscillations between machines. These harmonic waves cannot be filtered completely, thus the RoCoF measurement is not accurate anymore. Also, the estimation of system dynamics and the measurements of RoCoF require a distributed view of inertia. A robust method is proposed in this paper to determine the area suitable for frequency monitoring and RoCoF measurements, which maintains the accuracy of inertia estimation when oscillations exist.

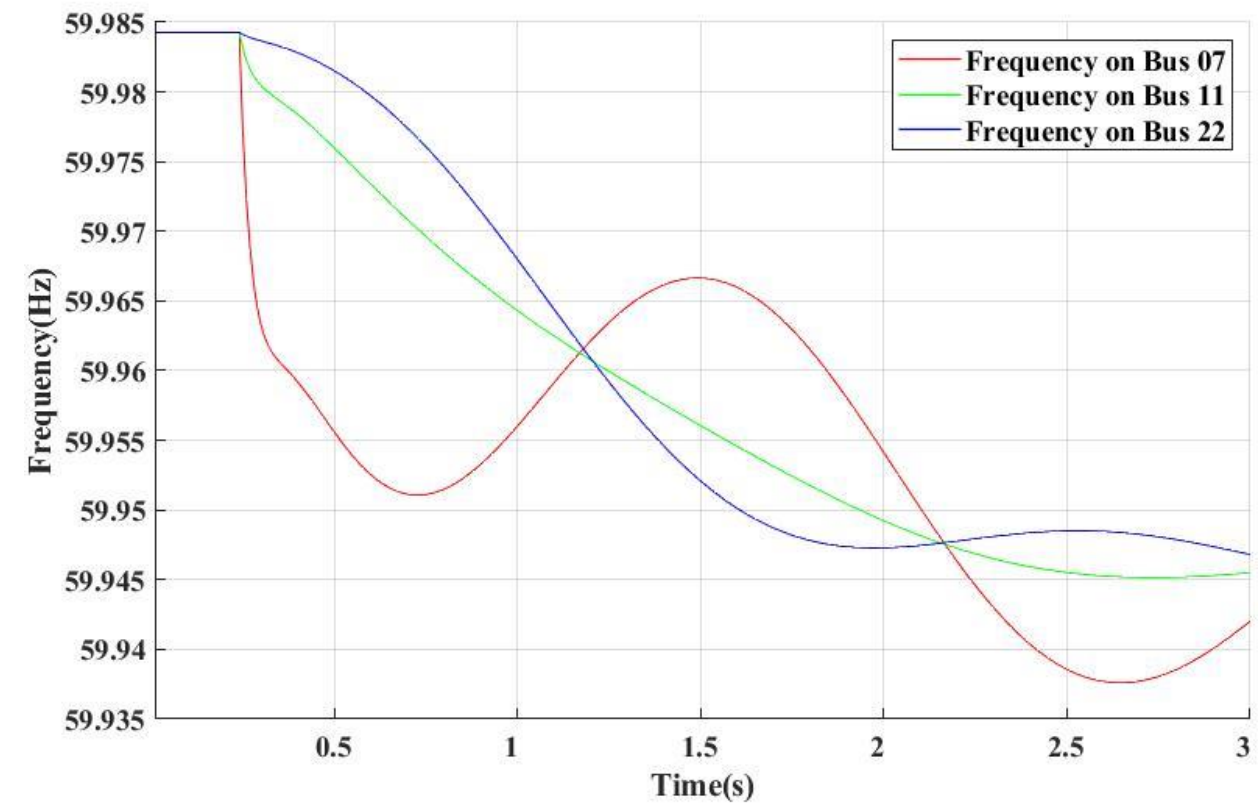

Fig. 2. System frequency trace after in-feed load on bus 8.

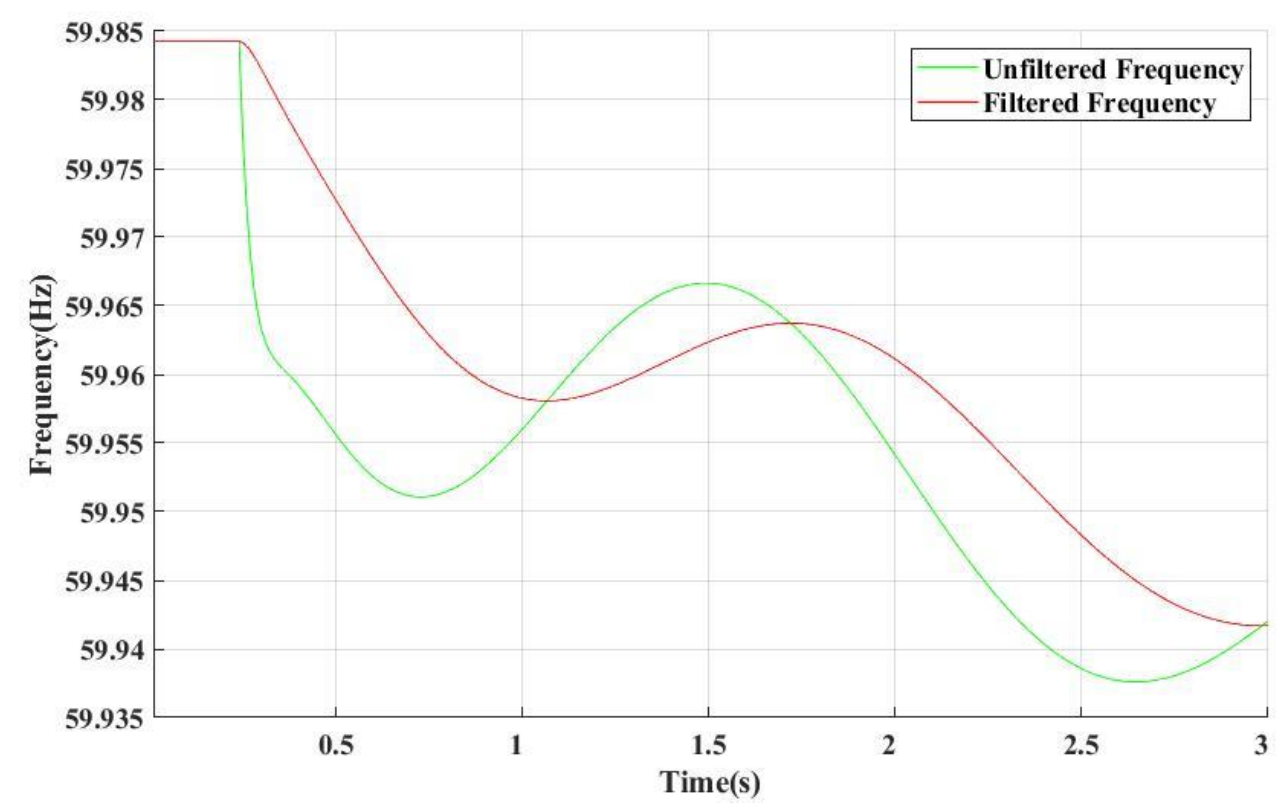

Fig. 3. Filtered frequency on bus 7 after in-feed load on bus 8.

### *B. Inertia Distribution Estimation*

However, for an equivalent COI model of a large system, the influence of power swing and oscillation dynamics cannot be neglected. The research presented in [17] demonstrated that a power system can be considered as multiple centers of inertia, coupled through the network. To locate the equivalent center of inertia and estimate the system inertia accurately, an inertia distribution index (IDI) is introduced, which has been proved to be highly linear correlation with system transfer function residue [12]. Availability of measurements from PMU makes it possible to evaluate the deviation of bus frequency from COI frequency in real time. The electrical distance from the monitored bus to COI location can be defined as follows,

$$dist(f_k, f_{COI}) = \int_{T_0+t_d}^{T+T_0+t_d} \left(f_k(\tau) - f_{COI}(\tau)\right)^2 d\tau \tag{9}$$

where $T_0$ is the time when a disturbance is detected, $t_d$ is the dead time considering the dead band of frequency, $T$ is the time length of the integration period to be determined, and $n$ is the total number of buses. Normalized inertia distribution index following a disturbance can be calculated as:

$$IDI_k = \frac{dist(f_k, f_{COI})}{\max\limits_{k\in\{1,..,n\}} dist(f_k, f_{COI})} \tag{10}$$

### C. Dynamic System Inertia Estimation

The value of $IDI_k$ reflects the electrical distance from bus $k$ to the COI location, the closest bus to the COI is determined as:

$$k_{COI} = \arg\min_{k\in\{1,..,n\}} IDI_k \qquad (11)$$

However, the COI location may not always be located at a particular bus; and $IDI_k$ may also vary under different disturbance events. Thus, to accurately estimate the system inertia, a clustering approach is proposed in this paper to determine the multi-bus COI area. Following a disturbance event, the bus $k_{COI}$ is selected as the initial mean of points in the COI area cluster, which represents that this bus is the most stable bus under the specific event. The electrical distance from an estimated bus to bus $k_{COI}$ can then be calculated below

$$dist(f_k, f_{k_{COI}}) = \int_{T_0+t_d}^{T+T_0+t_d} \left(f_k(\tau) - f_{k_{COI}}(\tau)\right)^2 d\tau \qquad (12)$$

where $f_{k_{COI}}$ is the measured frequency of the bus nearest to COI location.

The COI area $S_{COI}$ consists of buses that have electrical distances less than the pre-determined threshold value $\delta$.

$$S_{COI} = \{k: dist(f_k, f_{k_{COI}}) \le \delta\} \qquad (13)$$

It is known that the location of disturbance is a key factor in the inertia estimation. During a normal operation period, disturbance on different buses may cause distortion in bus frequency. To mitigate the impact of disturbance location on system inertia estimation, a dynamic COI area estimation method is proposed. During a specific time period, the system inertia is assumed to be stable under normal operation. We set a system observation window, within which events are detected while the system remains stable. The set of buses, $S_{COI}^T$, identified within the COI area over a period $T_{win}$ is defined in (14). The COI bus over period $T_{win}$, $k_{COI}^T$, is defined in (15).

$$S_{COI}^{T_{win}} = \left\{k^t: dist\left({f^t}_k, {f^t}_{k_{COI}}\right) \le \delta, t \in T_{win}\right\} \qquad (14)$$

$$k_{COI}^{T_{win}} = \arg\max_{k\in\{1,..,n\}} C_k \qquad (15)$$

where $C_k$ is the count of bus $k$ identified as a bus of the COI area over a period of $T_{win}$, $t$ indicates the event time within the observation window. Highest $C_k$ means bus $k$ is closest to the COI location and its changes in angle and frequency are minimal over period $T_{win}$. The impact of bus location on system inertia can be ranked by sorting $C_k$ in descending order. Here, $C_p$ is defined as the second highest index, which indicates that to some extent bus $p$ may represent the dynamics of system. When $C_p/C_k$ is larger than a threshold, which is 0.6 in this paper, it means the contribution from bus $p$ cannot be neglected; if $dist(f_k, f_{k_{COI}})$ also satisfies the criteria, the RoCoF $\mathrm{d}f_{est}/\mathrm{d}t$ then can be estimated as follows,

$$\frac{\mathrm{d}f_{est}}{\mathrm{d}t} = \frac{C_k}{C_p + C_k}\cdot\frac{\mathrm{d}f_k}{\mathrm{d}t} + \frac{C_p}{C_p + C_k}\cdot\frac{\mathrm{d}f_p}{\mathrm{d}t} \qquad (16)$$

where $\mathrm{d}f_k/\mathrm{d}t$ is the measured RoCoF on bus $k$, and $\mathrm{d}f_p/\mathrm{d}t$ is the measured RoCoF on bus $p$. If there is no feasible bus $p$, $C_p$ is set to 0.

The system inertia can be then estimated following the procedures shown in Fig. 4. The proposed dynamic inertia estimation method can effectively detect events and estimate the inertia accurately using the data extracted from WAMS system. If the size of loss is accurately known, then the total system inertia $E_{est}$ can be estimated:

$$E_{est} = \frac{f_0\Delta P}{2\frac{\mathrm{d}f_{est}}{\mathrm{d}t}} \qquad (17)$$

In a large system, inertia of a regional area can be estimated following a disturbance where the loss occurs outside the area; then $\Delta P$ can be extended to cover the total power crossing the area boundary

$$\Delta P = \sum_{i\in B} \Delta P_i \qquad (18)$$

where $\Delta P_i$ is the change in boundary exchange power in MW, $B$ is the set of boundary transmission lines.

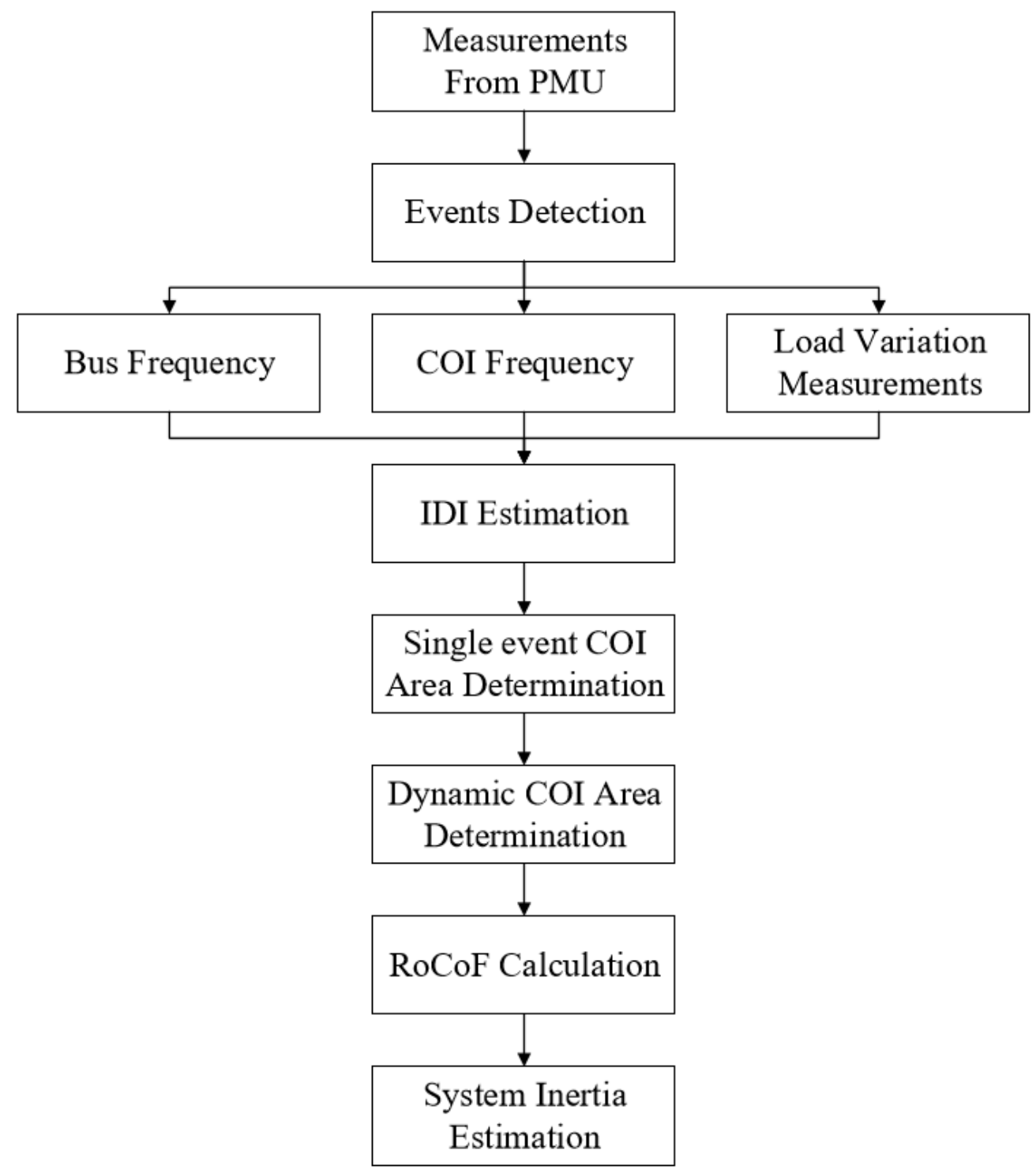

Fig. 4. System inertia estimation process on events.

## IV. Case Studies

The proposed approach is evaluated on the IEEE 24-bus test system. The system has 24 buses (17 buses with loads), 38 branches, 33 generators [18]. The total system load is 1,684 MW. The simulation model was implemented using TSAT, which is a core module of DSATools [19].

For an event where the disturbance appears as two distinct loss events, a non-monotonic frequency deviation may occur leading to erroneous IDI values. Integration period less than 0.5s can avoid the frequency distortion and make sure event detector captures more events. To determine the optimal integration period and ensure the IDI of the bus closest to COI location reaches the lowest value, sensitivity of integration period $T$ is tested on the base model. Fig. 5 shows the results of the sensitivity test of integration period. It can be observed that the IDI on bus 18, 21, 22 and 23 reach the lowest value at the integration period of 0.2 s. Thus, the integration period is set to 0.2 s.

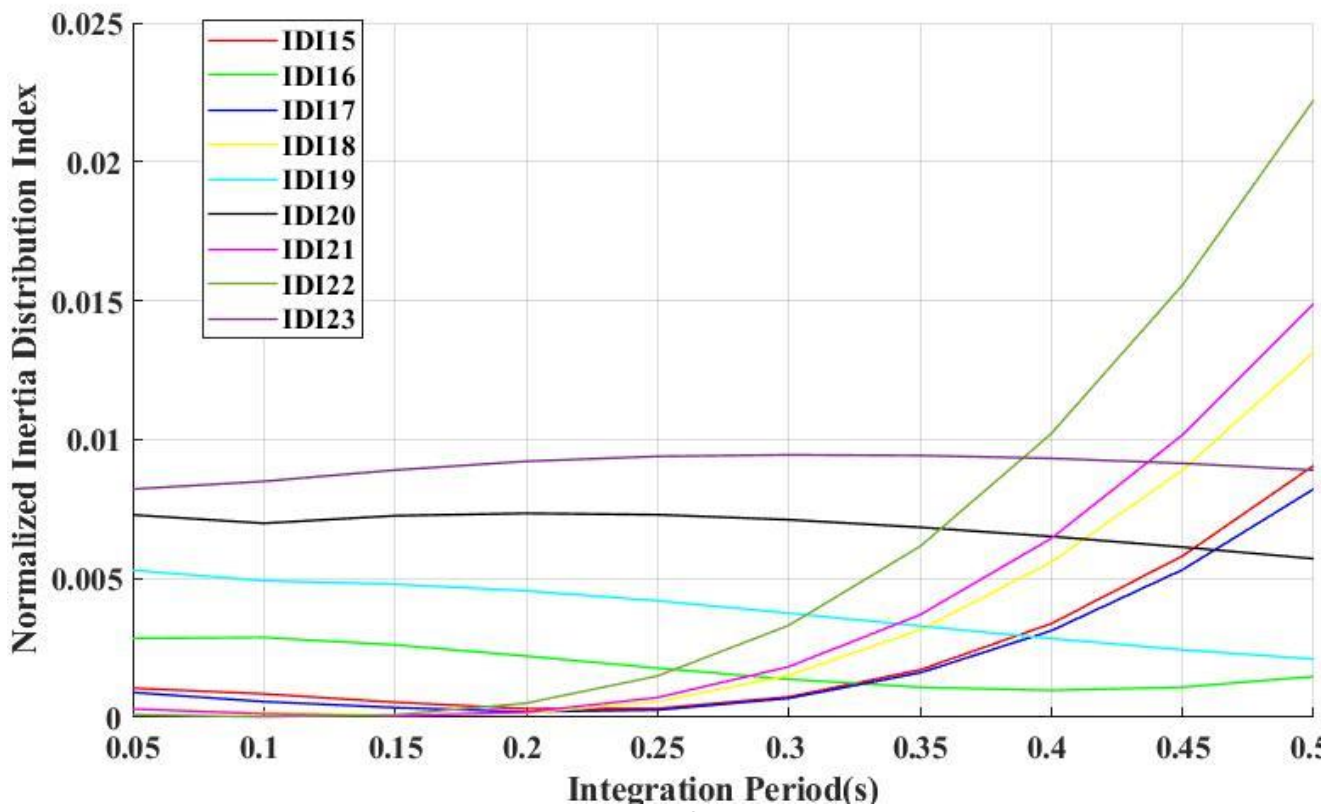

Fig. 5. Sensitivity of integration period.

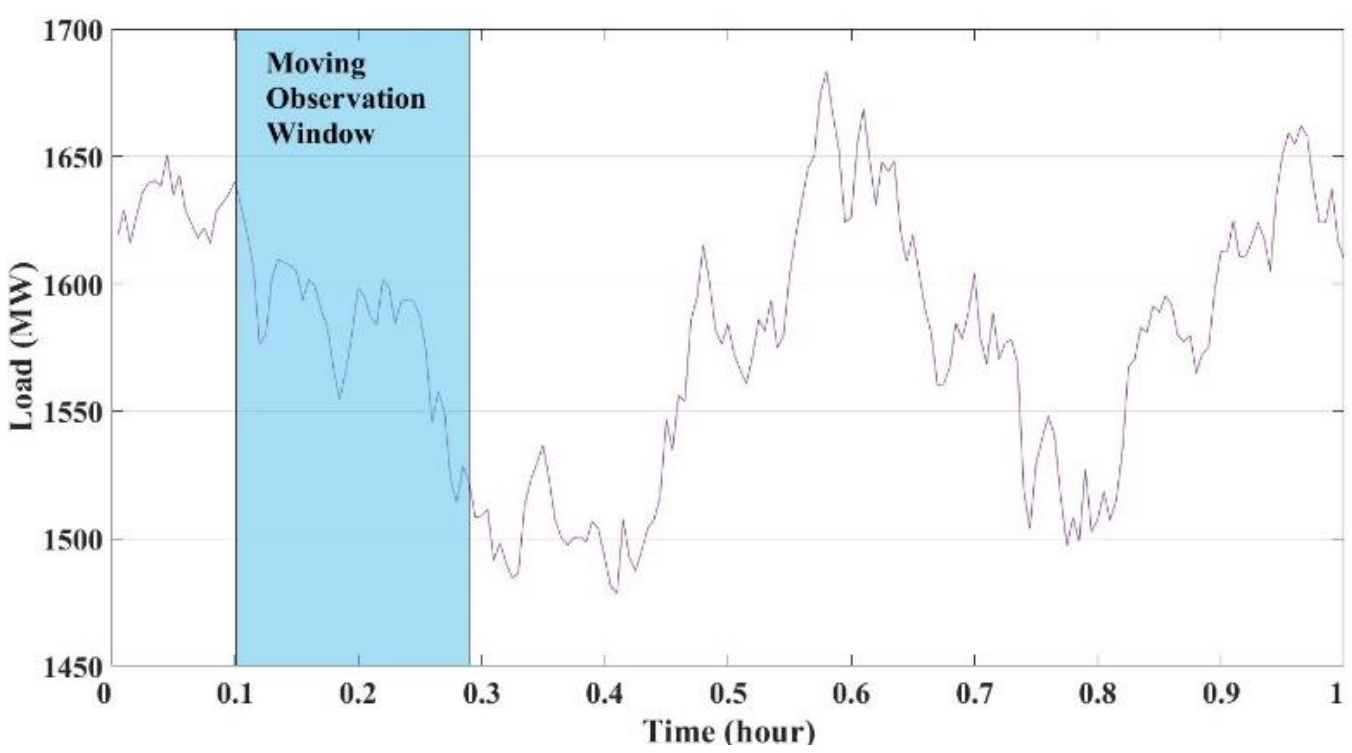

Fig. 6. Load variation profile.

The system total load profile is simulated for 60 minutes, as shown in Fig. 6. 100 perturbations occur evenly on each bus with same probability, and a moving 10-minute observation window is applied. In the first 10-minute window, 16 events are detected. Fig. 8 shows the result of the identified COI area buses in the 24-bus system using the proposed method, larger yellow circle means higher $C_k$ value of bus $k$. It is observed from Fig. 8 that bus 13 is identified as the COI bus of period $T_{win}$. While bus 23 also shows its close electrical distance to the COI location based on the proposed method. The results in Fig. 7 also show that frequency on bus 2, bus 7 and bus 22 contains harmonic waves. Inertia distribution index on these buses are estimated between 0.9 and 1, which indicates that these buses are relatively unstable and RoCoF measurements on these buses could suffer high bias.

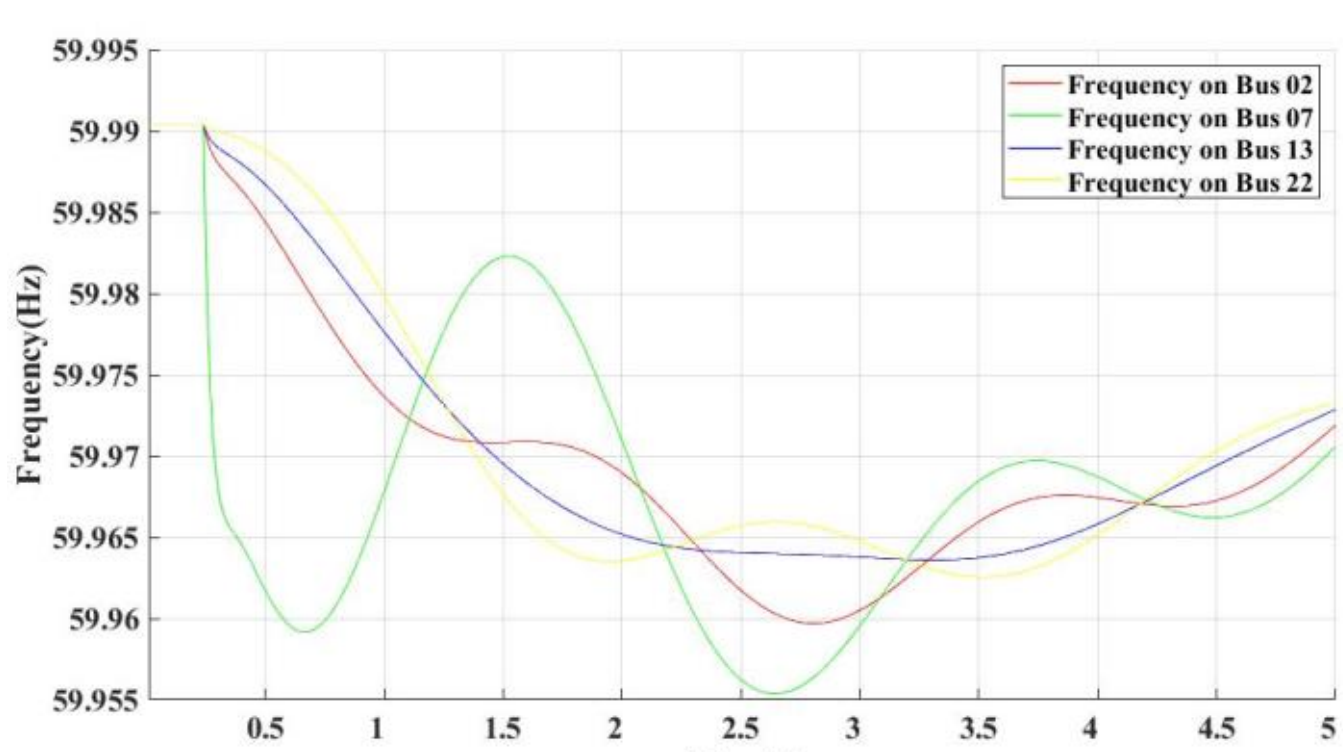

Fig. 7. Frequency measurements on different buses.

Method of curve fitting is used to mitigate the impacts of measured transients in frequency following a loss which leads to significant large RoCoF value. Fig. 9 shows the measured RoCoF on the determined COI bus, the measured RoCoF is corrected from -0.074 Hz/s to -0.046 Hz/s. The results of system inertia estimation, under a selected event, obtained with the proposed method are displayed in Table I. For a single detected event, the system inertia $H_{COI}$ is estimated as 30044.6 MWs$^2$ using the traditional method, while the real system inertia is 31,525 MWs$^2$; The corresponding percentage estimation error $\%H_{dif}^{COI}$ is -4.70%. The system inertia estimated with the proposed dynamic inertia estimation method, $H_{prop}$, is 30600.9 MWs$^2$ that corresponds to an estimation error $\%H_{dif}^{prop}$ of -2.93%, which shows substantial improvement over the results for traditional single event estimation method: the inertia estimation error dropped by 37.6% from 1480.4 MWs$^2$ to 924.1 MWs$^2$.This corresponds to an overall estimation improvement of 1.77%.

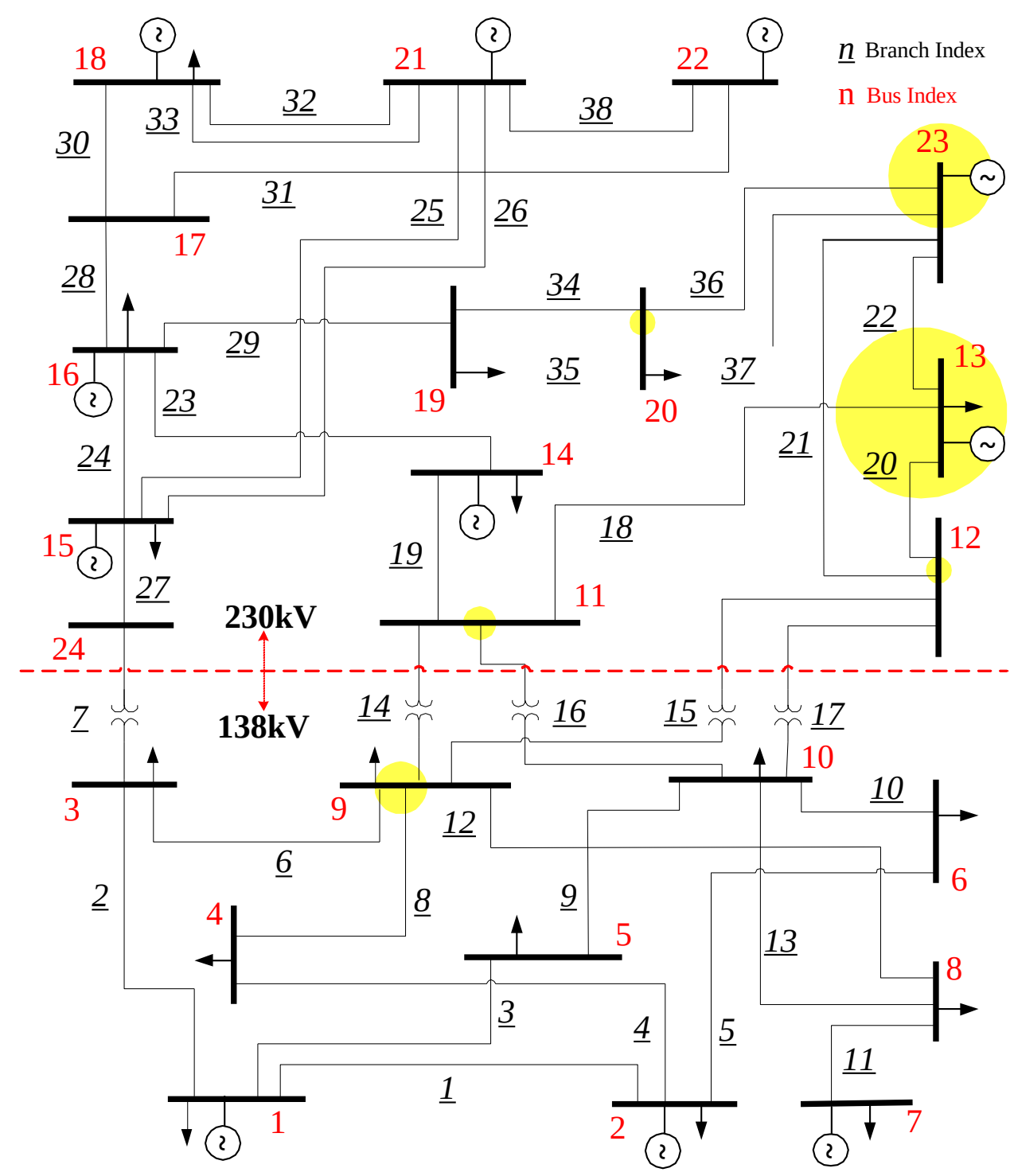

Fig. 8. Center of inertia area estimation [18].

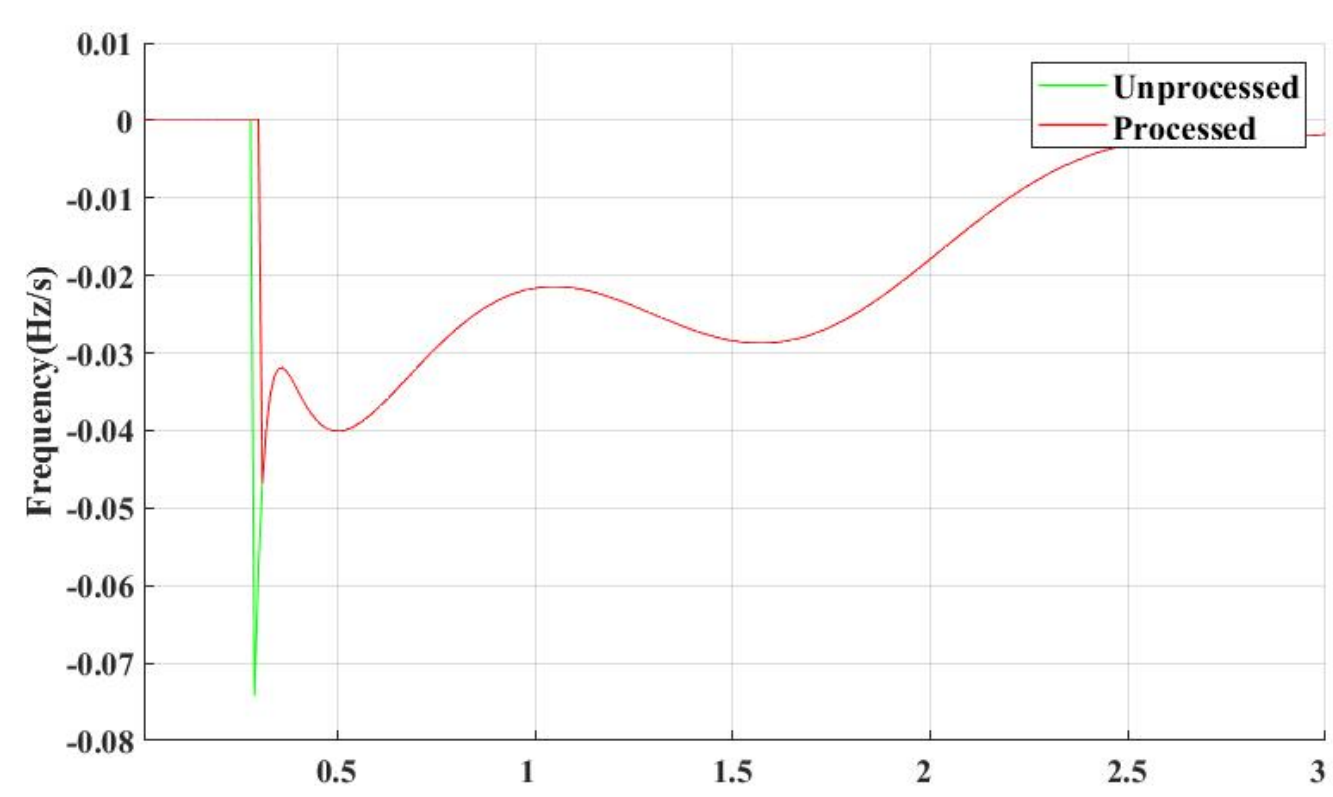

Fig. 9. RoCoF measurement in Center of inertia area.

Table I. Results of inertia estimation with various methods

| $\Delta P$ (MW) | $H_{real}$ ($MWs^2$) | $H_{COI}$ ($MWs^2$) | $\%H_{dif}^{COI}$ | $H_{prop}$ ($MWs^2$) | $\%H_{dif}^{prop}$ |
|---|---|---|---|---|---|
| 52.56 | 31525.0 | 30044.6 | -4.70% | 30600.9 | -2.93% |

To evaluate the impact of renewable penetration on inertia distribution, another emulation was conducted under scenario of 20% wind penetration level: generators on bus 2, bus 7 and bus 13 are replaced with wind generators. Fig. 10 shows the results under scenario of 20% wind penetration level. A significant excursion of COI location can be observed due to installation of wind plants. It shows that the COI location shifts towards the area where many synchronous generators are located and synchronized online.

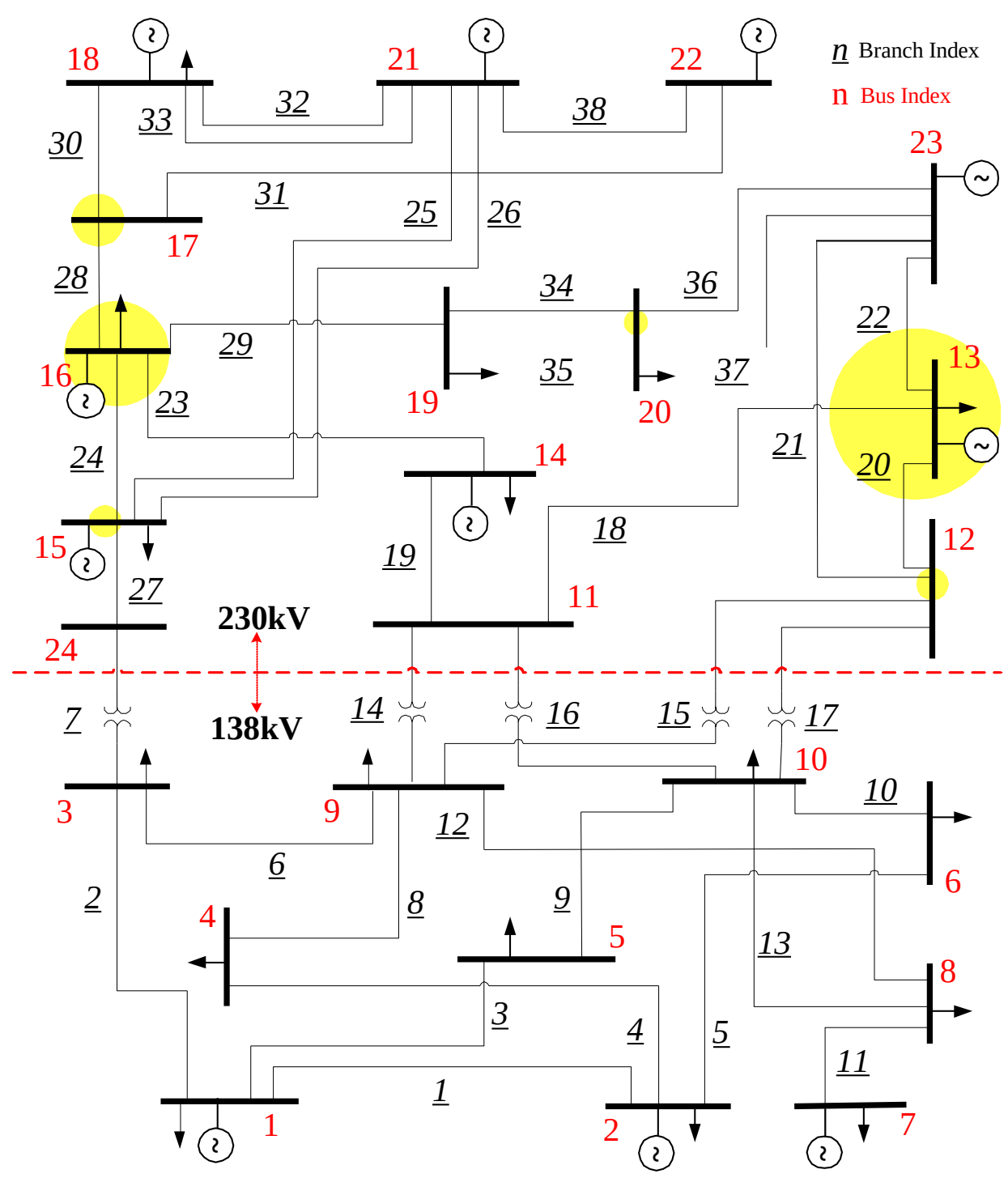

Fig. 10. Center of inertia area estimation with 20% wind generation penetration.

## V. CONCLUSIONS

As the integration of variable renewable generation increases, the reduction in the system inertia poses a serious challenge for frequency regulation. Evaluation of the system inertia distribution traditionally based on a single disturbance event may be susceptible to power swings and oscillation between machines, which could deteriorate the accuracy of measurements and lead to high biased estimation. Based on the equivalent center of inertia concept, a dynamic system inertia distribution estimation method is proposed in this paper.

The simulation results on the IEEE 24-bus system indicate that the power system with lower RES penetration shows a better frequency response, where the nadir is relatively higher and the RoCoF is less steep. The sensitivity test is then conducted to determine the optimal time length of integration period. The results also show that the proposed dynamic inertia estimation method utilizing the proposed clustering algorithm has a better performance on system inertia estimation by incorporating the impact of perturbation location and oscillation between machines. Buses within COI area show relative stability comparing to the neighbor areas, measurements on these buses are relative robust and authentic. Unstable buses, which suffer harmonic waves, are also determined during the estimation process. Finally, the impact of geographic location of RES on COI area is examined. Overall, the proposed method is more robust and accurate for estimating system inertia distribution. Potential applications using the concept of inertia distribution estimation would be explored in the future.